\documentclass[aps,preprint,superscriptaddress]{revtex4}
\usepackage{amsmath}
\usepackage{graphicx}

\begin{document}
\bibliographystyle{apsrev}


\title{Spin Degeneracy and Conductance Fluctuations in Open Quantum Dots }

\author{J. A. Folk, S. R. Patel, K. M. Birnbaum}
\affiliation{Department of Physics, Stanford University, Stanford,
California 94305}

\author{C. M. Marcus}
\affiliation{Department of Physics, Stanford University, Stanford,
California 94305}
\affiliation{Department of Physics, Harvard University, Cambridge, 
Massachusetts
02138}

\author{C. I. Duru\"oz and J. S. Harris, Jr.}
\affiliation{Department of Electrical Engineering, Stanford University,
Stanford, California 94305}

\begin{abstract} 
The dependence of conductance fluctuations
on parallel magnetic field is used as a probe of spin
degeneracy in open GaAs
quantum dots.
The variance of fluctuations at high parallel field is
reduced from the low-field variance (with
broken time-reversal symmetry) by factors ranging from roughly two in a $1\; \mu m^2$
dot to greater than four in $8\; \mu m^2$ dots.
The factor of two is expected
for Zeeman splitting of spin degenerate channels. A possible
explanation for the larger suppression based on field-dependent spin-orbit scattering is
proposed.
\end{abstract}

\maketitle

The combined influence of
coherence, confinement, electron-electron interactions, and spin makes the
understanding---not to mention the application---of quantum dots and other
coherent electronic structures extremely challenging.  For instance,
the role of interactions in breaking spin degeneracy in  confined systems
has been the subject of much recent work, both experimental \cite{Cobden98, Tans98,
Stewart97, Luescher00} and theoretical \cite{JacquodBrouwer00}, but no clear
concensus has emerged. In this Letter we explore the issue of spin
degeneracy in a regime lying between fully open
mesoscopic systems such as quantum point contacts, where spin
degeneracy is not broken (e.g., conductance plateaus occur at integer multiples
of $2e^2/h$  \cite{vanWees88}) and confined systems, where recent
experiments appear to show broken spin degeneracy  \cite{Tans98, Stewart97,
Thomas96, GaussianSpacings, Kouwenhoven97}.

Several recent experiments have probed the spin state of
discrete energy levels in nearly-isolated systems using Coulomb-blockaded
transport \cite{Cobden98, Tans98, Stewart97, Kouwenhoven97,
Ralph97}.  However, this approach typically
provides information about only a small sample of the level
spectrum and its spin structure, making results difficult to interpret in general
terms. For disordered or chaotic systems, it is often useful to take
a statistical approach to spectral and transport properties
\cite{AltshulerMeso, BeenakkerRMT,GuhrRMT}.  We will adopt this strategy, using the statistics
of conductance fluctuations
to investigate the degree of spin degeneracy in open quantum dots.

An open quantum dots is
connected to electron reservoirs via leads that pass one or more
fully transmitting channels. At low
temperatures, conductance through the dot fluctuates
randomly as a function of various external parameters such
as  perpendicular magnetic field or device shape.  Conductance
fluctuations in ballistic dots arise from the interference of
multiple transport paths through the
device, analogous to universal conductance fluctuations
observed in disordered mesoscopic samples (see Fig.\
1a).

The primary experimental signature that we investigate is the
reduction of these conductance fluctuations in large parallel
magnetic fields (see Figs.\ 1a and 1b).  Because conductance
fluctuations reflect spectral statistics, they are sensitive to both
time reversal symmetry---this aspect has been investigated in detail in previous work, Refs.
\cite{Chan,Huibers98,HuibersSwitkes98}---as well as spin degeneracy in the system.
To avoid the complicating effects of a parallel field on time-reversal symmetry,
measurements are reported {\em in all cases} with a small
perpendicular field applied (after confirming consistent behavior in these devices with previous
results \cite{Chan,Huibers98,HuibersSwitkes98}).

The original concept for the measurement was that if the system were spin degenerate
at low field, then a large in-plane field would lift the degeneracy via
Zeeman splitting, with associated changes in the amplitude of conductance
fluctuation. If, on the other hand, spin degeneracy at low field
 were already lifted by interactions, then a large parallel field would not alter
spectral statistics and hence conductance fluctuation amplitude. Surprisingly, we
find that the conductance fluctuations are indeed suppressed by a strong parallel
field (suggesting degeneracy at low field), but in many cases by a significantly
greater factor than can be understood in terms of a simple breaking of spin
degeneracy. At the end of the paper, we suggest a possible explanations for
this large supression as resulting from field-dependent spin-orbit scattering. This
possibility is fully explored theoretically in Ref.\ \cite{Halperin00}.

Breaking spin degeneracy changes the scattering matrix that describes linear
transport through the dot from the form
$\bigl(\begin{smallmatrix} a & 0 \\ 0 & a
\end{smallmatrix} \bigr) $  to the form
$\bigl(\begin{smallmatrix} a & 0 \\ 0 & b
\end{smallmatrix} \bigr)$, where $a$ and $b$
are scattering matrices
for the separate spin channels. Assuming independent random-matrix
statistics for $a$ and $b$ (appropriate for disordered or chaotic
systems), the variance of conductance fluctuations in dots is calculated to be two times larger for
spin-degenerate scattering matrices than for scattering matrices with broken spin degeneracy,
irrespective of the number of open channels
\cite{Lyanda94,Halperin00}.

A description  of mesoscopic fluctuations in terms of the statistics
of broadened energy levels of the dot provides an intuitive picture of how conductance
fluctuations can depend on
spin degeneracy \cite{Altshuler86}. At low
temperature, transport occurs coherently
through a number of levels proportional to the
escape rate,
$\Gamma_{esc}/\hbar$.  If levels corresponding to different spins do not
mix, only
levels of the same spin species will show level repulsion. In this
situation, each of the two spin species
contributes $\sim e^2/h$  to conductance fluctuations.  When the
spectrum is spin degenerate, 
fluctuations  from the two spin species add constructively, giving $var(g)
\sim (2e^2/h)^2$.
When spin degeneracy is broken, fluctuations instead add
randomly, giving $var(g)\sim 2(e^2/h)^2$.  Finite
temperature
reduces
$var(g)$ by a factor $kT/\Gamma_{esc}$, regardless of degeneracy \cite{Halperin00}.

Reduction of $var(g)$ due to magnetic-field-induced Zeeman splitting has been
observed previously in disordered mesoscopic systems.  In contrast
to the present observations, both
metal \cite{Moon96} and GaAs heterostructure \cite{Debray89} samples
show only the
factor of two reduction expected for spin-degenerate transport.

The quantum dots used in this experiment were fabricated on a
two-dimensional electron
gas (2DEG) formed at the interface of a GaAs/AlGaAs heterostructure
using Cr-Au surface
depletion gates pattered by standard electron beam lithography
 (Fig.\ 2
insets).  The GaAs/Al$_{0.3}$Ga$_{0.7}$As
interface was 40 $nm$ from the Si delta-doped layer ($n_{Si} =
1\times 10^{12}\;cm^{-3}$) and 90 $nm$ below
the wafer surface. The 2DEG density of $\sim 2 \times
10^{11}\;cm^{-2}$ and mobility
$\sim 1.4 \times 10^5\;cm^2/Vs$ gave a transport mean free path of
$\sim 1.5\;\mu m$.  Measurements were
made on three devices, one with area
$1\; \mu m^2$ and two with area $8\; \mu m^2$ (Figs.\ 2a,b
insets),
containing roughly $2\times 10^3$ and $1.6\times 10^4$ electrons respectively.
Standard 4-wire lock-in
techniques were used to measure conductance, with voltage
across the sample always less than
$kT/e$.  In all cases, noise was less than one tenth of conductance
fluctuation amplitude.  Measurements were performed in a dilution
refrigerator with a base mixing chamber temperature of $25\, mK$.  Electron
temperature of the reservoirs was measured independently using Coulomb
blockade peak width
\cite{KouwenhovenRev}, and was the same as that of the
mixing chamber over the range of temperatures reported.

The sample was oriented with the 2DEG parallel to the primary
magnetic field, as shown in Fig.\ 1c, in order to avoid
Landau quantization from the large fields required to
create significant Zeeman splitting ($\sim$ 1 to 7
$T$).  Smaller fields ($-0.1$
to $0.1$
$T$) were applied perpendicular to the 2DEG using an independent
split-coil magnet
that was attached to the outer vacuum can of the refrigerator (Fig.\
1d).  Slight sample misalignment ($<1^o$ from parallel) was determined by a
shift of the symmetry point in
$B_{perp}$ as a function of parallel field, and compensated by the
split-coil magnet.

Statistics of conductance fluctuations were gathered over
ensembles of dot shapes,
created by changing the voltages
applied to two gates, while the point
contacts were simultaneously adjusted to maintain constant
transmission.  At each parallel field, variance was
measured in several different
perpendicular fields, all shown together in Fig. 2.  An example of
conductance fluctuations as a function of two 
gate voltages is shown in Fig.\ 2c.  In the $1\;
\mu m^2$ dot, 450 shapes were sampled at each field, of which $\sim 200$ were
considered statistically independent; in the $8\; \mu m^2$ dot, 900 shapes
were sampled, of which $\sim 450$
were considered independent.  We emphasize that all ensembles were taken
with a perpendicular field sufficient to break time-reversal
symmetry in the devices.

The amplitude of conductance fluctuations was found to decrease and
then saturate upon
application of a parallel field of several tesla in all cases, as seen in
Fig.\ 2. In
most cases the reduction was
significantly larger than the expected factor of two (see Table in Fig.\
2).  In both
$8\;\mu m^2$ devices, $var(g)$ decreased by a factor of $\sim 4$ to  5; in the $1\;
\mu m^2$ dots, $var(g)$ decreased by a factor of $\sim 2$ at the
lowest temperatures and $\sim 3$ at higher
temperatures. The field scale for the reduction of
$var(g)$ increased with the number of channels in
the point contacts and with temperature (see Fig 2, table). Over the
same range of
parallel field, {\em average} conductances typically changed by
less than 5\%.

We are able to rule out the possibility that the
reduction in
$var(g)$ at high parallel field
was caused by either increased
temperature or increased dephasing, either of which would suppress conductance
fluctuations \cite{dephasingtheory, Huibers98}.
First, a direct measurement of electron gas temperature using Coulomb blockade
peak width indicated that for $ T \geq 100 mK$ a parallel field of
$4T$ increased electron temperature by
$\lesssim5\%$, relative to zero field.

To compare dephasing rates at low and high fields, we could not use
the standard measure
of dephasing---the magnitude of the weak localization correction to average
conductance---because fields larger than $\sim 0.5T$ were
observed to break time-reversal
symmetry even when strictly
parallel to the plane of the heterostructure.  Instead, dephasing
rates were compared using power
spectra of magnetoconductance
fluctuations, which for chaotic dots have the form
$S(f) \propto e^{-f/f_{0}}$, where $f$ is the frequency in cycles/mT, $f_{0}
\propto (N+\pi \hbar/(\Delta
\tau_\varphi))^{-1/2}$, $\Delta$ is the level spacing of the dot, and
$\tau_{\varphi}$ is the
dephasing time
\cite{Clarke95, Efetov95}. Note that the  characteristic frequency
$f_0$ has no explicit temperature dependence but does depend on
$\tau_{\varphi}$.  This
measure of dephasing rate has been shown to be consistent with weak
localization
measurements in quantum dots above $300 mK$
\cite{HuibersSwitkes98}.

Power spectra of conductance fluctuations at low and high parallel field, as
well as at higher temperature, are
shown in Fig.\ 3. All spectra clearly show the expected
$e^{-f/f_{0}}$ form, with the steeper slope for the
$250mK$ data showing that $f_0$ is indeed sensitive to dephasing.  From the
$4 T$ curve we observe that
$f_{0}$ is certainly not smaller than---and perhaps is even slightly
larger than---the
value at zero parallel field, suggesting that the dephasing rate
has not increased at large parallel field.

With time-reversal symmetry already broken,  orbital effects due to a parallel
field---including wave function compression or flux coupling due to a rough or asymmetric quantum well---should not
affect
$var(g)$. Having eliminated field-dependent temperature, decoherence, and orbital coupling as causes
of the reduced
$var(g)$, one is led to suspect that the effect may be
spin related.  Recalling the original motivation for the measurement, the
reduction in
$var(g)$ implies spin degeneracy at low field, up to an energy resolution
$\epsilon\sim max(\Gamma_{esc},kT)$, within the simple picture discussed above.
However, the fact that $var(g)$ is reduced by considerably more than the expected factor of two
with no increase in dephasing means that this simple picture must be incomplete.  Another difficulty
with the model of broken spin degeneracy is that the expected field scale for reducing
$var(g)$ should be given by
$g\mu B\sim\epsilon$. However, for the $8\;\mu m^2$ dot, where
$\Gamma_{esc}\ll kT$ for all temperatures measured, the field scale for the reduction was found not
to be proportional to temperature (see Table 1).

An interpretation of
the suppression of
$var(g)$ at high fields beyond a factor of two is that
there is a greater degree of spectral
rigidity than can be accounted for by Zeeman splitting of
spin-degenerate levels.  A mechanism that could
lead to this enhanced rigidity is spin-orbit scattering, which would
cause {\em all} levels in the spectrum
to repel. However, if spin-orbit scattering is significant in
explaining our results, its role must be
rather subtle.  First, the average conductance always shows weak
localization rather than anti-localization
around zero field over a broad range of temperatures and device
areas, indicating that
$\tau_{so}>\tau_{\varphi}$ at low fields. Second, if strong
spin-orbit scattering were present, the
perpendicular field necessary to break time-reversal symmetry
(present in all of these
measurements) would have been sufficient
to suppress $var(g)$ fully, and no further change would have been
observed as a function of
    parallel field. If, however, spin-orbit scattering increased upon application of a
    parallel field (leaving  spin-degeneracy intact at low field)  one
would expect a suppression in
$var(g)$ at high parallel field of greater than a factor of two while
still observing weak
localization (rather than
anti-localization) around zero field.  Interestingly, if a
field-dependent spin-orbit effect were
present, it would then  be the factor-of-{\em two} reduction of
$var(g)$ found in the  $1\;\mu m^2$ dot that
would become more difficult to explain. One needs, however, to
consider the dependence of these effects
on device size \cite{Halperin00}.

In conclusion, we have used conductance fluctuations to probe spin
degeneracy in open quantum dots.  We find that the
variance of the fluctuations is reduced at high parallel field,
implying that the low-field spectrum is
spin-degenerate to within $kT$ or escape broadening, $\Gamma_{esc}$.
The surprising observation
that $var(g)$ is reduced by significantly more than a factor of two in
certain cases has led us to consider
mechanisms other than the breaking of spin degeneracy that would
suppress $var(g)$, such as a field-dependent spin-orbit scattering rate.

We gratefully acknowledge discussions with A. Altland, B. Altshuler, I. Aleiner,
N. Birge, P. Brouwer, J. Cremers, L. Glazman, B.
Halperin,  Y. Oreg, B. Simons, and A. Stern.  This work was supported in part
by the ARO under 341-6091-1-MOD 1 and DAAD19-99-1-0252.  JAF
acknowledges support from the DoD.
\vspace{0.1in}

\begin{thebibliography}{10}

\expandafter\ifx\csname bibnamefont\endcsname\relax
     \def\bibnamefont#1{#1}\fi
\expandafter\ifx\csname bibfnamefont\endcsname\relax
     \def\bibfnamefont#1{#1}\fi
\expandafter\ifx\csname url\endcsname\relax
     \def\url#1{\texttt{#1}}\fi
\expandafter\ifx\csname urlprefix\endcsname\relax\def\urlprefix{URL }\fi
\expandafter\ifx\csname bibinfo\endcsname\relax \def\bibinfo#1#2{#2}\fi
\expandafter\ifx\csname eprint\endcsname\relax \def\eprint#1{#1}\fi
\bibitem{Cobden98}
\bibinfo{author}{\bibfnamefont{D.~H.} \bibnamefont{Cobden}} \emph{et~al.},
     \bibinfo{journal}{Phys. Rev. Lett.}
     \textbf{\bibinfo{volume}{81}}, \bibinfo{pages}{681}
     (\bibinfo{year}{1998}).

\bibitem{Tans98}
\bibinfo{author}{\bibfnamefont{S.}~\bibnamefont{Tans}},
     \bibinfo{author}{\bibfnamefont{M.~H.} \bibnamefont{Devoret}},
     \bibinfo{author}{\bibfnamefont{R.~J.~A.} \bibnamefont{Groeneveld}},
     \bibnamefont{and} \bibinfo{author}{\bibfnamefont{C.}~\bibnamefont{Dekker}},
     \bibinfo{journal}{Nature}
     \textbf{\bibinfo{volume}{394}}, \bibinfo{pages}{761}
     (\bibinfo{year}{1998}).

\bibitem{Stewart97}
\bibinfo{author}{\bibfnamefont{D.~R.} \bibnamefont{Stewart}} \emph{et~al.},
     \bibinfo{journal}{Science}
     \textbf{\bibinfo{volume}{278}}, \bibinfo{pages}{1784}
     (\bibinfo{year}{1997});
\bibinfo{author}{\bibfnamefont{D.}~\bibnamefont{Stewart}},
     \emph{\bibinfo{title}{Level Spectroscopy in Quantum Dots}}, Ph.D. thesis,
     \bibinfo{school}{Stanford University} (\bibinfo{year}{1998}).

\bibitem{Luescher00} S. Luescher \emph{et~al.}, cond-mat/0002226 (2000).

\bibitem{JacquodBrouwer00}
\bibinfo{author}{\bibfnamefont{P.~W.} \bibnamefont{Brouwer}},
     \bibinfo{author}{\bibfnamefont{Y.}~\bibnamefont{Oreg}}, \bibnamefont{and}
     \bibinfo{author}{\bibfnamefont{B.~I.} \bibnamefont{Halperin}},
     \bibinfo{journal}{Phys. Rev. B} \textbf{\bibinfo{volume}{60}},
     \bibinfo{pages}{13977} (\bibinfo{year}{2000});
\bibinfo{author}{\bibfnamefont{P.}~\bibnamefont{Jacquod}}
\bibnamefont{and}
     \bibinfo{author}{\bibfnamefont{A.~D.} \bibnamefont{Stone}},
     \bibinfo{journal}{Phys. Rev. Lett.} \textbf{\bibinfo{volume}{84}},
     \bibinfo{pages}{3938} (\bibinfo{year}{2000});
  H. U. Baranger, D.~
Ullmo, L.~I.~Glazman, Phys. Rev. B {\bf 61},
R2425 (2000); I.~L.~Kurland, I.~L.~Aleiner, and B.~L.~Altshuler,
cond-mat/0004205 (2000).

\bibitem{vanWees88}
\bibinfo{author}{\bibfnamefont{B.~J.} \bibnamefont{van Wees}} \emph{et~al.},
     \bibinfo{journal}{Phys. Rev. Lett.}
     \textbf{\bibinfo{volume}{60}}, \bibinfo{pages}{848}
     (\bibinfo{year}{1988}).

\bibitem{Thomas96}
\bibinfo{author}{\bibfnamefont{K.~J.} \bibnamefont{Thomas}} \emph{et~al.},
     \bibinfo{journal}{Phys. Rev. Lett.}
     \textbf{\bibinfo{volume}{77}}, \bibinfo{pages}{135}
     (\bibinfo{year}{1996}).



\bibitem{GaussianSpacings}
\bibinfo{author}{\bibfnamefont{U.} \bibnamefont{Sivan}} \emph{et~al.}
\bibinfo{journal}{Phys. Rev.
Lett.}
     \textbf{\bibinfo{volume}{77}}, \bibinfo{pages}{1123}
     (\bibinfo{year}{1996});
\bibinfo{author}{\bibfnamefont{S.~R.} \bibnamefont{Patel}}
\emph{et~al.},  \bibinfo{journal}{Phys. Rev.
Lett.}
     \textbf{\bibinfo{volume}{80}}, \bibinfo{pages}{4522}
     (\bibinfo{year}{1998});
\bibinfo{author}{\bibfnamefont{F.} \bibnamefont{Simmel}} \emph{et~al.}
\bibinfo{journal}{Phys. Rev.
B}
     \textbf{\bibinfo{volume}{59}}, \bibinfo{pages}{10441}
     (\bibinfo{year}{1999}).

\bibitem{Kouwenhoven97}
\bibinfo{author}{\bibfnamefont{L.~P.} \bibnamefont{Kouwenhoven}} \emph{et~al.},
     \bibinfo{journal}{Science}
     \textbf{\bibinfo{volume}{278}}, \bibinfo{pages}{1788}
     (\bibinfo{year}{1997}).


\bibitem{Ralph97}
\bibinfo{author}{\bibfnamefont{D.~C.} \bibnamefont{Ralph}},
     \bibinfo{author}{\bibfnamefont{C.~T.} \bibnamefont{Black}},
\bibnamefont{and}
     \bibinfo{author}{\bibfnamefont{M.}~\bibnamefont{Tinkham}},
     \bibinfo{journal}{Phys. Rev. Lett.}
     \textbf{\bibinfo{volume}{78}}, \bibinfo{pages}{4087}
     (\bibinfo{year}{1997}).

\bibitem{AltshulerMeso}
\bibinfo{editor}{\bibfnamefont{B.~L.}~\bibnamefont{Altshuler}},
     \bibinfo{editor}{\bibfnamefont{P.~A.} \bibnamefont{Lee}}, \bibnamefont{and}
     \bibinfo{editor}{\bibfnamefont{R.}~\bibnamefont{Webb}}, eds.,
     \emph{\bibinfo{title}{Mesoscopic Phenomena in Solids}}
     (\bibinfo{publisher}{North-Holland}, \bibinfo{address}{Elsevier},
     \bibinfo{year}{1991}).

\bibitem{BeenakkerRMT}
\bibinfo{author}{\bibfnamefont{C.~W.~J.} \bibnamefont{Beenakker}},
     \bibinfo{journal}{Rev. Mod. Phys.} \textbf{\bibinfo{volume}{69}},
     \bibinfo{pages}{731} (\bibinfo{year}{1997}).

\bibitem{GuhrRMT}
\bibinfo{author}{\bibfnamefont{T.}~\bibnamefont{Guhr}},
     \bibinfo{author}{\bibfnamefont{A.}~\bibnamefont{M\"{u}ller-Groeling}},
     \bibnamefont{and} \bibinfo{author}{\bibfnamefont{H.~A.}
     \bibnamefont{Weidenm\"{u}ller}}, \bibinfo{journal}{Phys. Rept.}
     \textbf{\bibinfo{volume}{299}}, \bibinfo{pages}{189} 
(\bibinfo{year}{1998}).

\bibitem{Chan}
\bibinfo{author}{\bibfnamefont{I.~H.} \bibnamefont{Chan}}
\emph{et~al.},
     \bibinfo{journal}{Phys. Rev. Lett.}
     \textbf{\bibinfo{volume}{74}}, \bibinfo{pages}{3876}
     (\bibinfo{year}{1995}).

\bibitem{Huibers98}
\bibinfo{author}{\bibfnamefont{A.~G.} \bibnamefont{Huibers}} \emph{et~al.},
     \bibinfo{journal}{Phys. Rev. Lett.}
     \textbf{\bibinfo{volume}{81}}, \bibinfo{pages}{1917}
     (\bibinfo{year}{1998}).

\bibitem{HuibersSwitkes98}
\bibinfo{author}{\bibfnamefont{A.~G.} \bibnamefont{Huibers}} \emph{et~al.},
     \bibinfo{journal}{Phys. Rev. Lett.}
     \textbf{\bibinfo{volume}{81}}, \bibinfo{pages}{200}
     (\bibinfo{year}{1998}).

\bibitem{Halperin00} B.~I.~Halperin  \emph{et~al.}, cond-mat/0010064 (2000).

\bibitem{Lyanda94}
\bibinfo{author}{\bibfnamefont{Y.~B.} \bibnamefont{Lyanda-Geller}}
     \bibnamefont{and} \bibinfo{author}{\bibfnamefont{A.~D.}
     \bibnamefont{Mirlin}}, \bibinfo{journal}{Phys. Rev. Lett.}
     \textbf{\bibinfo{volume}{72}}, \bibinfo{pages}{1894}
     (\bibinfo{year}{1994}).

\bibitem{Altshuler86}
\bibinfo{author}{\bibfnamefont{B.~L.} \bibnamefont{Altshuler}},
     \bibnamefont{and} \bibinfo{author}{\bibfnamefont{B.~I.}
\bibnamefont{Shklovskii}},
     \bibinfo{journal}{Soviet Physics - JETP}
     \textbf{\bibinfo{volume}{64}}, \bibinfo{pages}{127}
     (\bibinfo{year}{1986}).

\bibitem{Moon96}
\bibinfo{author}{\bibfnamefont{J.~S.} \bibnamefont{Moon}},
     \bibinfo{author}{\bibfnamefont{N.~O.} \bibnamefont{Birge}},
\bibnamefont{and}
     \bibinfo{author}{\bibfnamefont{B.}~\bibnamefont{Golding}},
     \bibinfo{journal}{Phys. Rev. B}
     \textbf{\bibinfo{volume}{53}}, \bibinfo{pages}{R4193}
     (\bibinfo{year}{1996});
\bibinfo{author}{\bibfnamefont{J.~S.} \bibnamefont{Moon}},
     \bibinfo{author}{\bibfnamefont{N.~O.} \bibnamefont{Birge}},
\bibnamefont{and}
     \bibinfo{author}{\bibfnamefont{B.}~\bibnamefont{Golding}},
     \bibinfo{journal}{Phys. Rev. B}
     \textbf{\bibinfo{volume}{56}}, \bibinfo{pages}{15124}
     (\bibinfo{year}{1997}).

\bibitem{Debray89}
\bibinfo{author}{\bibfnamefont{P.}~\bibnamefont{Debray}},
     \bibinfo{author}{\bibfnamefont{J.-L.} \bibnamefont{Pichard}},
     \bibinfo{author}{\bibfnamefont{J.}~\bibnamefont{Vicente}}, 
\bibnamefont{and}
     \bibinfo{author}{\bibfnamefont{P.~N.} \bibnamefont{Tung}},
     \bibinfo{journal}{Phys. Rev. Lett.}
     \textbf{\bibinfo{volume}{63}}, \bibinfo{pages}{2264}
     (\bibinfo{year}{1989}).

\bibitem{KouwenhovenRev}
\bibinfo{author}{\bibfnamefont{L.~P.} \bibnamefont{Kouwenhoven}} \emph{et~al.},
     in \emph{\bibinfo{booktitle}{Mesoscopic Electron Transport}}, edited by
     \bibinfo{editor}{\bibfnamefont{L.~L.}~\bibnamefont{Sohn}},
     \bibinfo{editor}{\bibfnamefont{L.~P.}~\bibnamefont{Kouwenhoven}},
     \bibnamefont{and}
\bibinfo{editor}{\bibfnamefont{G.}~\bibnamefont{Sch\"{o}n}}
     (\bibinfo{publisher}{Kluwer}, \bibinfo{address}{Dordecht},
     \bibinfo{year}{1997}).

\bibitem{dephasingtheory}
\bibinfo{author}{\bibfnamefont{H.~U.}~\bibnamefont{Baranger}} \bibnamefont{and}
     \bibinfo{author}{\bibfnamefont{P.~A.}~\bibnamefont{Mello}},
     \bibinfo{journal}{Phys. Rev. B}
     \textbf{\bibinfo{volume}{51}}, \bibinfo{pages}{4703}
     (\bibinfo{year}{1995});
\bibinfo{author}{\bibfnamefont{P.~W.}~\bibnamefont{Brouwer}} \bibnamefont{and}
     \bibinfo{author}{\bibfnamefont{C.~W.~J.}~\bibnamefont{Beenakker}},
     \bibinfo{journal}{Phys. Rev. B}
     \textbf{\bibinfo{volume}{51}}, \bibinfo{pages}{7739}
     (\bibinfo{year}{1995});
I.~L.~Aleiner and A.~I.~Larkin, Phys. Rev. B {\bf 54}, 14423 (1996);
\bibinfo{author}{\bibfnamefont{P.~W.}~\bibnamefont{Brouwer}} \bibnamefont{and}
     \bibinfo{author}{\bibfnamefont{C.~W.~J.}~\bibnamefont{Beenakker}},
     \bibinfo{journal}{Phys. Rev. B}
     \textbf{\bibinfo{volume}{55}}, \bibinfo{pages}{4695}
     (\bibinfo{year}{1997}).



\bibitem{Clarke95}
\bibinfo{author}{\bibfnamefont{R.~M.} \bibnamefont{Clarke}} \emph{et~al.},
     \bibinfo{journal}{Phys. Rev. B}
     \textbf{\bibinfo{volume}{52}}, \bibinfo{pages}{2656}
     (\bibinfo{year}{1995}).

\bibitem{Efetov95}
\bibinfo{author}{\bibfnamefont{K.~B.} \bibnamefont{Efetov}},
     \bibinfo{journal}{Phys. Rev. Lett.}
     \textbf{\bibinfo{volume}{74}}, \bibinfo{pages}{2299}
     (\bibinfo{year}{1995}).


\end{thebibliography}
\footnotesize{

}

\begin{figure}[bt] \centering
    \label{fig1}
    \includegraphics[width=5in]{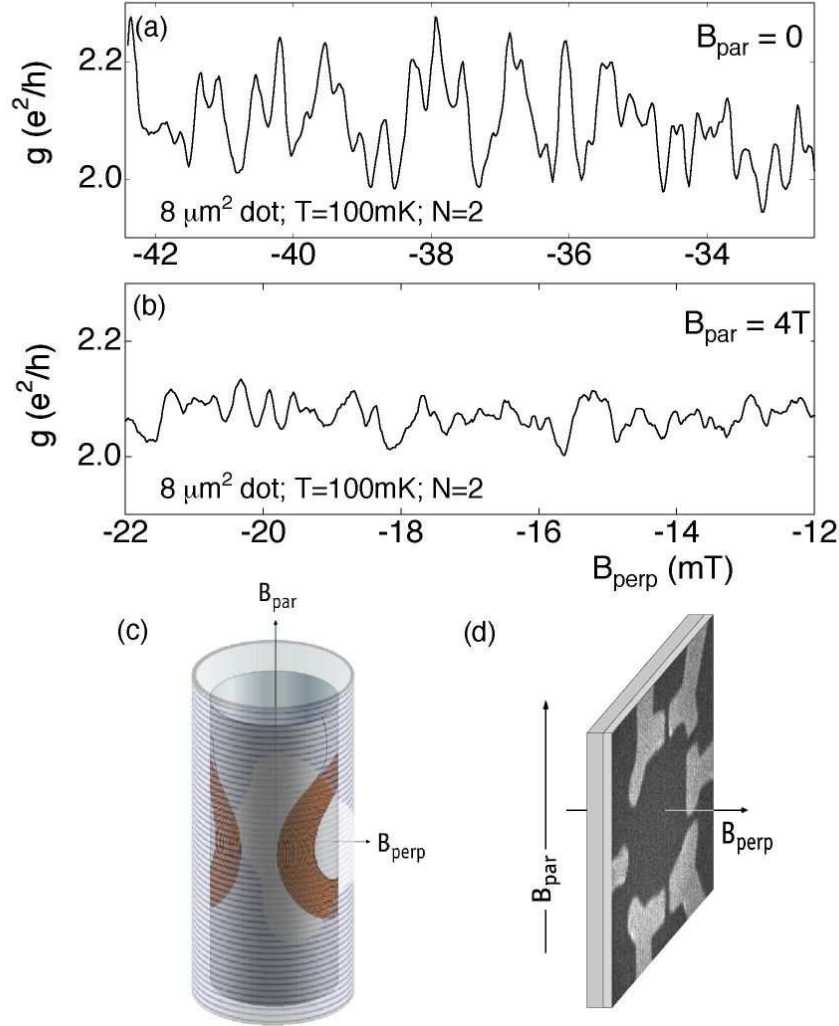}
    \caption{\footnotesize {A sample of conductance fluctuations in one of the $8\, \mu m^2$
 dots, as a function of perpendicular field, $B_{perp}$, at (a) zero parallel field,
$B_{par}=0$, and (b) $B_{par}=4T$.  Horizontal axes represent field applied through perpendicular
coils only; different ranges compensate for small perpendicular component of $4T$ field in (b),
and thus represent the same actual perpendicular field.   (c)
Illustration showing placement of superconducting coils used to generate
$B_{perp}$ relative to the vacuum can and the primary solenoid used to produce
$B_{par}$. (d) Schematic indicating
orientation of
$B_{par}$ and $B_{perp}$ with respect to the planar quantum dot (The orientation of
$B_{par}$ within the plane is not accurately depicted and has not been investigated.)}}
    \end{figure}

 \begin{figure}[bt]\centering
    \label{fig2}
    \includegraphics[width=5in]{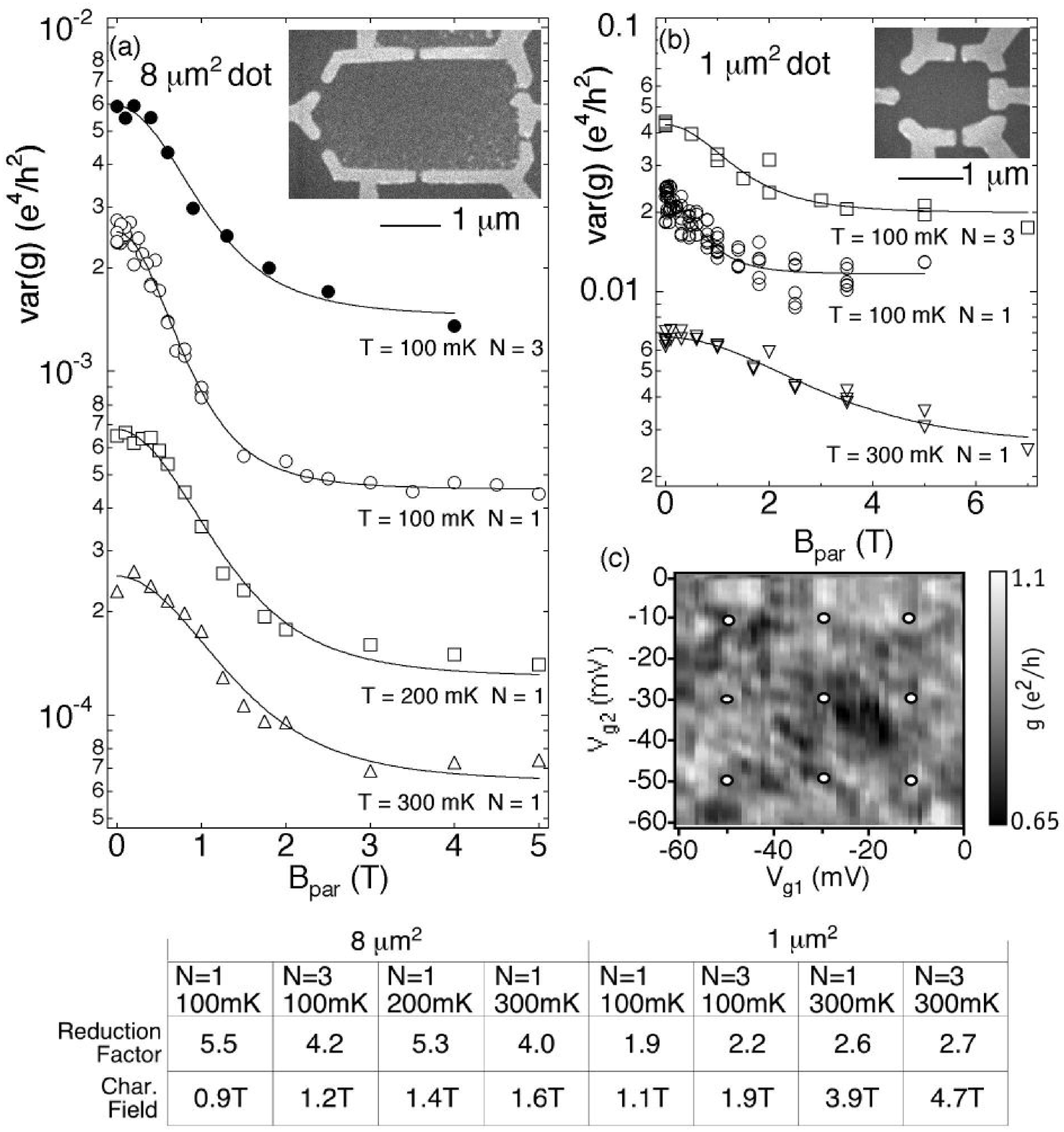}
    \caption{\footnotesize{Variance of conductance fluctuations,  $var(g)$, as a
function of parallel field, $B_{par}$, for
(a) an
$8\,
\mu m^2$ dot and (b) a
$1\, \mu m^2$ dot, at several temperatures, $T$, and numbers of
   modes, $N$, in each lead. Fits to a Lorentzian-squared form (solid curves)
were used to extract the magnitude and characteristic field for the
reduction in $var(g)$.  Changes in the magnitude of
$var(g)$ as a function of temperature reflect thermal averaging and
dephasing. (c) Conductance (greyscale) versus  shape-distorting gate
voltages
$V_{g1}$ and
$V_{g2}$ in an $8\, \mu m^2$ device, showing sampling in shape space (white
dots) relative to characteristic scale of fluctuations. (Several hundred
samples are used to find each value of $var(g)$; see text.)  Table:
Reduction factors of
$var(g)$, and characteristic parallel fields, from  Lorentzian-squared
fits.}}
    \end{figure}

  \begin{figure}[bt]\centering
    \label{fig3}
    \includegraphics[width=4in]{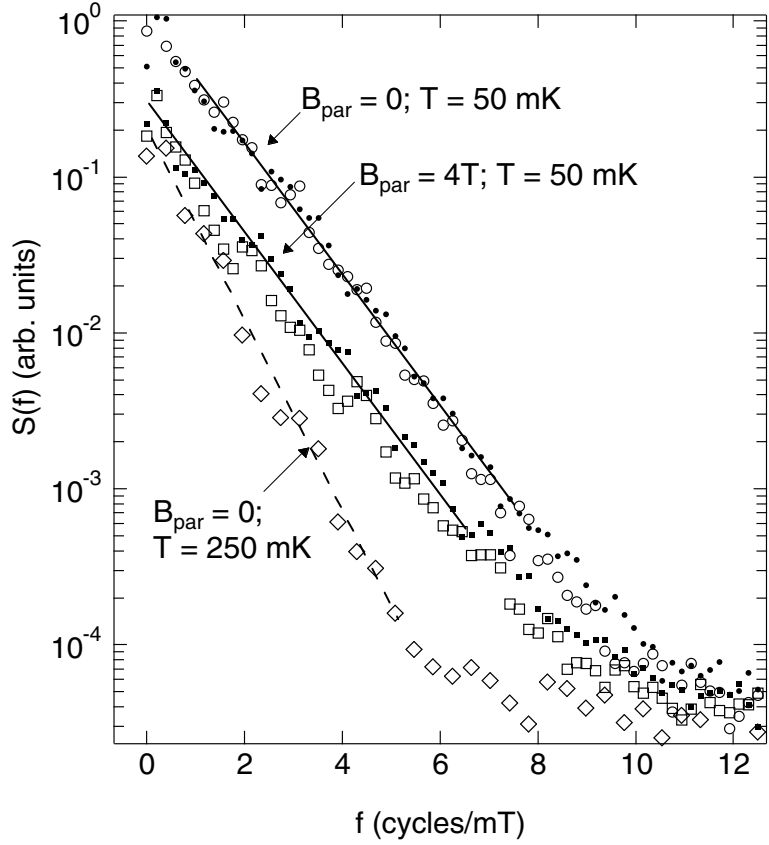}
    \caption{\footnotesize{Power spectra $S(f)$ of conductance
fluctuations in an
$8\; \mu m^2$ dot at $B_{par}=0$ for
$50mK$ (circles) and $250mK$ (diamonds), and at
$B_{par}=4T$ for $50mK$ (squares).  Open and filled markers show
different shape ensembles.
$S(f)$ has the expected exponential form (see text), with
characteristic frequency $f_0$ given by the slope
in the log-linear plot.  Identical slopes at $B_{par}=0$ and $4T$
(parallel solid lines)
indicate no change in dephasing at high field.  The steeper slope of
$S(f)$ at $250mK$ (dashed
line) indicates increased dephasing.}}
    \end{figure}

\end{document}